\documentclass[a4paper,11pt]{article}
\usepackage{natbib}
\usepackage{amsmath}
\usepackage{amssymb}
\usepackage{graphicx}

\newcommand{\HMM}{hidden Markov model}
\newcommand{\anHMM}{a hidden Markov model}
\newcommand{\HMMs}{hidden Markov models}
\newcommand{\ucHMMs}{Hidden Markov models}
\newcommand{\MCMC}{Markov chain Monte Carlo}
\newcommand{\anMCMC}{a Markov chain Monte Carlo}
\newcommand{\cTc}{{\cal{T}}_c}
\newcommand{\Xspace}{\mathbb{X}}
\newcommand{\Sspace}{\mathbb{S}}
\newcommand{\oldT}{{\cal{T}}}
\newcommand{\oldX}{{\cal{X}}}
\newcommand{\newT}{{\cal{T}'}}
\newcommand{\newX}{{\cal{X}'}}
\newcommand{\fixT}{\overline{\cal{T}}}
\newcommand{\fixX}{\overline{\cal{X}}}
\newcommand{\Xobs}{{\cal{X}_\text{obs}}}
\newcommand{\bs}{\mathbf{s}}
\newcommand{\bX}{X}
\newcommand{\bx}{x}

\begin{document}
\title{\textbf{Integrated Continuous-time\\ Hidden Markov Models}}
\author{Paul G. Blackwell}
\date{\textit{School of Mathematics and Statistics,\\ University of Sheffield, Sheffield S3 7RH, U.K.\\ p.blackwell@sheffield.ac.uk}\\[2ex]\today} 
\maketitle
\begin{abstract}
\noindent
Motivated by applications in movement ecology, in this paper I propose a new class of \textit{integrated continuous-time hidden Markov models} in which each observation depends on the underlying state of the process over the whole interval since the previous observation, not only on its current state. This class gives a new representation of a range of existing models, including some widely applied switching diffusion models. I show that under appropriate conditioning, a model in this class can be regarded as a conventional hidden Markov model, enabling use of the Forward Algorithm for efficient evaluation of its likelihood without sampling of its state sequence. This leads to an algorithm for inference which is more efficient, and scales better with the amount of data, than existing methods. This is demonstrated and quantified in some applications to animal movement data and some related simulation experiments.
\end{abstract}


\section{Introduction}

The motivation for this paper comes from the analysis of animal movement data, arising for example from GPS tagging. This type of application has received a great deal of attention recently; see for example the review by \citet{Toby}. Typically the animal's location $X(t)$ is observed at discrete, sometimes regular, instants in time $t_1,t_2,\ldots$. Conceptualisation of the process often involves an underlying behavioural state $S(t)$, with the movement process switching between different forms depending on that behaviour. 
\ucHMMs\ (defined in \S\ref{standardHMM}) have thus been widely used to model movement, with the `hidden' state representing the behaviour \citep{Morales2004}. However, the application of this approach in continuous-time modelling \cite[e.g.][]{Blackwell1997,Exact} has been limited by computational complexity, as the existing algorithms for \HMMs\ do not immediately apply; see \citet{Toby} for discussion. The aim of this paper is to show how to carry out fast, exact computation, with an algorithm closely related to the Forward Algorithm of a conventional \HMM\ (see \S\ref{standardHMM}), for a broad class of continuous-time models, including many suitable for representing animal movement. Its use is illustrated in some real examples using telemetry data, and its performance compared with existing methods.

\section{Existing classes of model}

\subsection{Hidden Markov Models}

\label{standardHMM}

A \HMM\ is an unobserved discrete-time Markov chain $\{S_k\}$ accompanied by observations $Y_k$, with the distribution of each observation determined by the corresponding value of the chain, so that 
\[
Y_j \sim f_{S_j}(\cdot).
\]
In the simplest case, the observations are conditionally independent given the chain. In a movement context, the `observation' in this sense is some function of the sequence of observed locations, for example 
the displacement $Y_j = X(t_{j})-X(t_{j-1})$.

\ucHMMs\ are very widely studied and applied, and it is beyond the scope of the current paper to fully review modelling with \HMMs\ and methods for inference. However, one key factor in their wide adoption is the existence of a highly efficient algorithm, the Forward Algorithm, allowing the calculation of the probability of a sequence of observations by indirectly summing over all possible state sequences, a calculation that would be prohibitively expensive if carried out na\"ively. See for example \cite{ZucchiniMacDonald} for general background, and \cite{moveHMM} and \cite{Toby} for specific applications to discrete-time models of animal movement.

\subsection{Continuous-time Hidden Markov Models}
\label{sec:cthmm}

A continuous-time Hidden Markov Model is an unobserved continuous-time Markov chain $\{S(t)\}$ accompanied by conditionally independent observations $Y_{t_1}, Y_{t_2}, \ldots$ with distributions determined by the values of the chain at a countable set of times, so that 
\[
Y_{t_i} \sim f_{S(t_i)}(\cdot).
\]
For examples in medical contexts, see \citet{JacksonSharples2002, Liu2017}.
This case can be handled using broadly the same inferential methods as a standard \HMM, with the form of the transition matrices dependent on the time intervals $t_i-t_{i-1}$, and subject to some constraints even if the observations are equally spaced (since not every discrete-time Markov chain can be expressed as a restriction of a time-homogeneous continuous-time Markov chain to equally spaced times). This model has what is known as the `snapshot' property: the distribution of the observation $Y_{t_i}$ depends only on the state at the same instant, $S(t_i)$, and conditional on $S(t_i)$ is independent of $S(t), t \not=t_i$.

Thinking about movement in continuous time, an animal's location $X(t_i)$ naturally depends on its behaviour between $t_{i-1}$ and $t_i$; that is, $X(t_i)$ depends both on $X(t_{i-1})$ and on the whole of $\{S(t), t_{i-1}<t<t_i \}$, as discussed by \citet[\S4.4]{Toby}. The process does not have the `snapshot' property and as a consequence, cannot be represented as a continuous-time Hidden Markov Model in the sense defined above. The computational approach, in particular the Forward Algorithm, that gives such power to the usual \HMM\ does not immediately apply.

\section{Integrated continuous-time Hidden Markov Models}

For more flexible modelling, it is useful to consider a Markov process $Z(t)=(X(t),S(t))$ on $\Xspace\times\Sspace$, where $\Sspace$ is discrete. If either (a) $\Xspace$ is continuous and the regularity conditions given by \cite{Berman} for such processes are satisfied, or (b) $\Xspace$ is discrete, then $S(t)$ is piecewise constant over time, with transition rates $\lambda_{ij}(t, x),~i,j\in \Sspace$ say.

I define an \textit{integrated continuous-time \HMM} (InCH) to be a Markov process $Z(t)$ as above, satisfying one of the conditions (a) and (b), with rates $\lambda_{ij}(t, x)$ that are bounded. 
In general, it lacks the snapshot property defined in \S\ref{sec:cthmm}, since the way $X(t)$ is changing depends on $S(t)$. This class includes a wide range of existing models; the reason for formulating them in this particular way is the potential improvement in computational efficiency permitted if these conditions are met, as described in \S\ref{InCH}.

In a movement context, usually $\Xspace$ will be continuous. In particular, the separable switching diffusion models of \cite{Exact} can be thought of as InCH models on $\mathbb{R}^d\times\{1,\ldots,n\}$. Usually we will be interested in $d=2$, but cases with $d=1$ and $d=3$ arise naturally. Related applications involving multiple animals lead to higher values of $d$; see \citet{Mu}.

It is useful to distinguish some particular cases. An InCH is \textit{spatially homogeneous} if $\lambda_{ij}(t, x)$ is independent of $x$; otherwise it is \textit{spatially heterogeneous}. Of course, $X(t)$ need not represent geographical space, but the terminology is appropriate to many applications, and makes the necessary distinction from time-homogeneity. 

In the next section, I show how to carry out computation using ideas closely related to the conventional \HMM\ or the `snapshot' case, for both spatially heterogeneous and spatially homogeneous InCH models. While the applications are certainly not limited to animal movement, it is convenient to use the terms `location' and `behaviour' to refer to the components of a process $Z$. Similarly, while models with discrete $\Xspace$ are certainly possible, the particular interest here is in continuous $\Xspace$, and I will refer to the density of  $X$, for simplicity.

\section{Representation and Algorithms}

\subsection{Uniformization}
\label{uniform}

Consider an InCH process with transition rates $\lambda_{ij}(t, x)$. Let $\lambda_{i}(t, x) = \sum_{j\not=i}\lambda_{ij}(t, x)$ represent the rate of switching out of behaviour $i$, at time $t$, when at location $x$, and 
let $\kappa$ be an upper bound so that $\kappa \geq \lambda_{i}(t, x) ~\forall i,t,x$.
Then, following \cite{Exact}, the occurrences of changes in behaviour can be represented as a dynamic thinning of a Poisson process of potential switches of uniform rate $\kappa$, with retention probability
$\lambda_{S(t)}(t, X(t))/\kappa$.
The unthinned Poisson process of potential switches does not depend on $t$ and $x$, enabling us to partially separate location and behaviour in a way that turns out to be crucial for inference.

If the $\lambda_{ij}(t, \bx)$ are known, for example if we are interested purely in simulating a known process, then  we can simply take $\kappa = \sup_{i,t,\bx}\{\lambda_{i}(t, \bx)\}$. 
In the more general inference context, the most straightforward case, for both exposition and implementation, is when the prior support of $\lambda_{ij}(t, \bx)$ is bounded above by $u_{ij}(t, \bx)$, with the function $u_{ij}(t, \bx)$ also bounded above. We can then define $u_{i}(t, \bx) = \sum_{j\neq i}u_{ij}(t, \bx)$
and take $\kappa = \sup_{i,t,\bx}\{u_{i}(t, \bx)\}$. This is the approach taken in \S\ref{sec:ibex} and \S\ref{sec:kinkajou}. If the priors are not all bounded above, then $\kappa$ is not fixed, and must be sampled in the inference process; see the discussion in \S\ref{discuss}.

In the spatially homogeneous case, the behaviour process can be thought of as a Markov chain on $\Sspace$ subordinated to a $\text{Poisson}(\kappa)$ process; that is, a continuous-time Markov chain in which some `events' do not change the state of the process.

\citet{RaoTeh} make use of the idea of uniformization, in the context of inference for a continuous-time Markov chain, and give additional background on the concept, including a proof (in the time-homogeneous case) of the representation described above.

\subsection{Existing inference methods}
For discrete-time \HMMs, `snapshot' continuous-time \HMMs\ and continuous-time Markov chains, a range of efficient algorithms for inference are available, as already indicated. Here, I focus on existing methods specific to switching diffusions and similar models, for comparison with the new methods in \S\ref{InCH} below.

\citet{Exact} make use of the uniformization representation in \S\ref{uniform}, with a Markov chain Monte Carlo 
algorithm that relies on forward simulation of the model between potential switches, sampling not only the times of the potential switches but also the locations and states at those times. This enables exact inference, in the sense that there is no time discretisation or approximation, and so the limiting distribution for the chain is exactly the joint posterior distribution of trajectories and parameters. It is widely applicable because of the flexibility in specifying the transition rates. However, because the state space for the \MCMC\ algorithm includes the behaviour not only at the observation times but also at the unknown collection of potential switching times, the algorithm is computationally demanding and mixes relatively slowly. 

In the spatially heterogeneous case, more efficient updates that do not rely purely on forward simulation are possible, by proposing locations and states in a more general way. For example, it is possible to propose a reconstruction of part of the behaviour sequence without reference to the locations, from a spatially homogeneous continuous-time Markov chain, conditioning only on the behaviour at the start and end of the interval being updated, and then propose corresponding locations, given the behaviour, from the movement process conditioned to give the appropriate time-inhomogeneous bridge; the acceptance probability then accounts for the difference between the true behaviour process and the proposal distribution. Alternatively, locations can be proposed at potential switching times using some time-homogeneous bridge process, and then behaviours proposed from the true behaviour process given the locations; again, the acceptance probability can account for the difference between the true and proposal movement processes. 
Experimentation suggests that these algorithms can be worthwhile in particular cases; however, for most purposes they are likely to be superseded by the algorithms introduced in \S\ref{InCH} below.

In the spatially homogeneous case of \cite{Blackwell1997}, where transition rates do not depend on location, behavioural trajectories $S(\cdot)$ can be sampled within \anMCMC\ algorithm without sampling the locations $\bX(\cdot)$ associated with the transitions. This kind of algorithm, as detailed in \cite{Blackwell2003}, does not use uniformization, nor would it particularly benefit from it. However, the algorithms of the next section offer great benefits in efficiency in this homogeneous case too, and do rely on uniformization, combined with some additional simplification that exploits the homogeneity.

\subsection{Inference for integrated continuous-time hidden Markov models}
\label{InCH}
A much more efficient inferential approach can be developed by exploiting the fact that, conditional on the times and locations corresponding to potential switches, an InCH process is effectively a time-inhomogeneous version of a conventional discrete-time \HMM, defined at the potential switching times, in which the transition probabilities are 
\begin{align*}
p_{ij}(t,\bx) &= \lambda_{ij}(t,\bx)/\kappa, ~ i\neq j\\
p_{ii}(t,\bx) &= 1-\lambda_{i}(t,\bx)/\kappa,
\end{align*}
and the observations are given by the changes in location, with
\[
f(\bX(t_{k+1})\mid t_k, \bX(t_{k}), S(t_{k}))
\]
given by the density of the movement process corresponding to behaviour $S(t_{k})$, 
\[
f_{S(t_{k})} (\bX(t_{k+1})\mid t_k, \bX(t_{k})).
\]
This differs from a typical \HMM\ in that the transition probabilities and observation densities are highly variable between time points. Nevertheless, the standard Forward Algorithm that enables evaluation of the likelihood for \anHMM\ without the need for explicit sampling of the states still applies here, and will enable the calculation of the likelihood for this model very efficiently. 

Of course, the times of potential switches are \emph{not} known, and the corresponding locations are \emph{not} observed. Instead, we observe the locations, and typically not the behavioural states, at a set of known times which may or may not be regularly spaced. Thus in practice we need to embed the evaluation of the likelihood using the Forward Algorithm within \anMCMC\ algorithm; but that algorithm now has a state space of \emph{much} lower dimension, not involving the behavioural states which are now integrated out. 

The method of \citet{RaoTeh} has some similarities. They are primarily interested in the continuous-time Markov chain (or Markov Jump Process, in their terminology) itself, and do not integrate it out; instead they use the Forward Algorithm to permit fast updating of a part of the realisation of the chain. 

The details of the InCH approach are given in the two following sections, which deal separately with the spatially heterogeneous and homogeneous cases. The former is easier to describe, and so is given first; 
it covers a rich class of models, allowing general spatio-temporal covariates in movement modelling, for example. The latter represents an important special case in applications, and it permits the integration out of locations, for further improvement in efficiency where applicable.

\subsection{Spatially heterogeneous case}
\label{sec:hetero}
Here we need to consider \anHMM\ defined at a set of times which is the union of the potential switching times, at which the transition matrix is defined as above in \S\ref{InCH}, and the observation times, at which (with probability 1) no change in state occurs and so the transition matrix is just the identity matrix $I_n$. Spatial locations at the potential switching times need to be sampled within \anMCMC\ algorithm, but the states do not.

Write $t_c$ for the $c$th observation time, and $t_{c,k}$ for the $k$th potential switching time between $t_c$ and $t_{c+1}$, for $k=1,\ldots,M_c$. Of course, for any given $c$, $M_c$ may be zero. 

An outline of the key step in the new \MCMC\ algorithm is as follows.
Choose $a,b$ such that $1\leq a<b \leq n_\text{obs}$. 
Define $\oldT = \{t_{c,k}| c=a,\ldots,b-1, k=1,\ldots,M_c\}$ and 
$\oldX = \{\bX(t_{c,k}) | c=a,\ldots,b-1, k=1,\ldots,M_c\}$.
Propose new counts $M'_c, c=a,\ldots,b-1$, times $\newT = \{t'_{c,k} | c=a,\ldots,b-1, k=1,\ldots,M'_c\}$
and locations $\newX = \{\bX'(t'_{c,k}) | c=a,\ldots,b-1, k=1,\ldots,M'_c\}$.
Evaluate the Hastings ratio based on likelihoods that integrate over all state sequences, replacing 
$\oldT,\oldX$ with $\newT,\newX$, 
and accept or reject accordingly.

We could choose new values $\newX$ independently of $\oldX$, for simplicity, or close to $\oldX$ to allow `small' steps that retain the information in $\oldX$. That is, we could take either an independence sampling or a random walk approach. For maximum flexibility, we formulate the proposals as a mixture of these two extremes. 

In more detail, we propose $t'_{c,k}$ and $\bX'(t'_{c,k})$ for a given $c$ as follows. $M'_c$ is proposed independently of $M_c$, with $M'_c \sim \text{Poisson}((t_{c+1}-t_c)\kappa)$, and $t'_{c,k}$ are defined as the order statistics of $M'_c$ independent uniform random variables on $(t_{c},t_{c+1})$. 
We define $\mu_\text{I}, \Sigma_\text{I}$ as the mean and covariance respectively of a Brownian bridge with diffusion parameter $\omega$ from $\bX(t_c)$ to $\bX(t_{c+1})$, evaluated at times $t'_{c,k}, k=1,\ldots,M'_c$, corresponding to the idea of an independent proposal. Similarly we write $\mu_\text{D}, \Sigma_\text{D}$ for the mean and covariance of a series of Brownian bridges with diffusion parameter $\omega$ passing through $\bX(t_c), \bX(t_{c,1}),\ldots,\bX(t_{c,{M_c}}), \bX(t_{c+1})$, again evaluated at $t'_{c,k}, k=1,\ldots,M'_c$. We propose new locations for a given $c$ from a weighted mixture of these bridges, 
\begin{align}
\label{mvn}
\bX'(t'_{c,1}),\ldots,\bX'(t'_{c,{M'_c}}) \sim \text{Normal}(p\mu_\text{I} + (1-p)\mu_\text{D}, p^2\Sigma_\text{I}+(1-p)^2\Sigma_\text{D}).
\end{align}
The proposal for $\newX$ is just the collection of these proposals for $c=a,\ldots,b-1$.
Both $\omega$ and $p$ are effectively tuning parameters. 

To calculate the Hastings ratio, define $\Theta$ to be all model and tuning parameters plus the initial distribution of the state, $\fixT$ to be all the observation times plus the potential switches outside $(t_a, t_b)$, and $\fixX$ to be the corresponding locations.
Then
\begin{align*}
p(\oldX,\oldT\mid \fixX, \fixT, \Theta) &= p(\oldX,\oldT,\fixX \mid \fixT, \Theta)/p(\fixX\mid \fixT, \Theta)\\
&\propto p(\fixX,\oldX\mid \fixT,\oldT,\Theta)p(\oldT\mid \Theta),
\end{align*}
where the first term is exactly the probability given by running the Forward Algorithm for $\fixX,\oldX$, and the second just defines a Poisson process of rate $\kappa$ on $(t_a,t_b)$.
Writing $q(\cdot)$ for a generic proposal density, we also have
\[
q(\newX,\newT\mid\oldX,\oldT,\fixX,\fixT,\Theta) = q(\newT\mid\Theta)q(\newX\mid\oldT,\fixT,\oldX,\fixX,\Theta),
\]
where the first term again defines a Poisson$(\kappa)$ process on $(t_a,t_b)$, and the second is the product over $c$ of densities of multivariate normal distributions of the form in equation (\ref{mvn}).
The Hastings ratio $\text(HR)(\newT,\newX\mid\oldT,\oldX)$ is then 
\begin{align*}
& \frac{q(\oldX,\oldT\mid\newX,\newT,\fixX,\fixT,\Theta)}{q(\newX,\newT\mid\oldX,\oldT,\fixX,\fixT,\Theta)} \times
\frac{p(\newX,\newT\mid \fixX, \fixT, \Theta)}{p(\oldX,\oldT\mid \fixX, \fixT, \Theta)}\\
=~&
\frac{q(\oldX\mid\newT,\fixT,\newX,\fixX,\Theta)}{q(\newX\mid\oldT,\fixT,\oldX,\fixX,\Theta)} \times
\frac{p(\fixX,\newX\mid \fixT,\newT,\Theta)}{p(\fixX,\oldX\mid \fixT,\oldT,\Theta)},
\end{align*}
since the terms from the Poisson process cancel, and is easily calculated usng the Forward Algorithm and equation (1). 

The other steps in the overall \MCMC\ algorithm are straightforward, sampling the model parameters by using random walk proposals and evaluating the likelihood using the Forward Algorithm.

\subsection{Spatially homogeneous case}
\label{SpatHom}
In the special case where the InCH process is spatially homogeneous \emph{and} the movement processes are solutions to a linear stochastic differential equation, we can completely avoid the need to sample the locations $x(t_{c,k})$, integrating them out using a matrix calculation which can be thought of as a special case of Kalman Filtering. 

As discussed at length elsewhere (\cite{Blackwell1997,Blackwell2003,Exact}), models where the movement process for each state is defined by a linear stochastic differential equation can lead to surprisingly rich behaviour, so this case is of practical importance. In particular, even the case where movement simply switches between different speeds of Brownian motion is important in data analysis; see \citet{Kranstauber2012} and the example in \S\ref{BM}.
For spatially homogeneous but non-linear models, one straightforward option would be to apply the `heterogeneous' methods above, but the example from \citet{AlisonJABES} of continuous-time step-and-turn models suggests that a more efficient compromise ought to be possible.
Such cases are not explicitly considered in this paper; for the remainder of this section, I concentrate on the homogeneous linear case.

Spatial homogeneity means that the transition probabilities of the uniform chain do not depend on the locations at the potential switching times, and linearity implies that for any given sequence of behaviours, movement densities can be calculated explicitly even over time intervals that incorporate changes in behaviour.
For particular states $i$ and $j$ at times $t_c$ and $t_{c+1}$, with potential switching times 
\[
\cTc = t_{c,1},\ldots,t_{c,M_c}
\]
we have
\begin{equation}
f_{ij}(\bx(t_{c+1})\mid\bx(t_{c}), \cTc) 
= \sum_{\bs} \pi_{ij}(\bs) \phi_{ij}(\bx(t_{c+1})\mid\bx(t_{c}),\cTc,\bs)
\label{SpatHomMeth}
\end{equation}
where 
\[
\bs = (s_1,\ldots,s_{M_c-1})
\]
is a possible sequence of states entered at times $t_{c,1},\ldots,t_{c,M_c-1}$, 
\[
\pi_{ij}(\bs) = p_{i,s_1}\times\cdots\times p_{s_{M_c-1},j},
\]
each $p_{i,j}$ is a transition probability as derived in \S\ref{InCH}, and $\phi_{ij}(\cdot\mid\cdot)$ is the transition density conditional on the sequence and timing of states. In general, $\phi_{ij}(\cdot\mid\cdot)$ can be calculated as a density from a $d$-dimensional normal distribution with parameters calculated recursively as in \S3.3 of \cite{Blackwell2003}. See \S\ref{BM} below for a special case.

This summation over sequences of states is, of course, exactly the kind of calculation that the \HMM\ Forward Algorithm is designed to avoid, because its computational cost increases rapidly with the number of time points. As a brute-force way of calculating the likelihood globally, it would be impractical because it scales so badly with the size of the data-set. As used here however, for calculation of the likelihood locally between successive observations, it is feasible provided $\kappa$ is not too large, which will be true in cases where the data are reasonably informative about the model.

Given this explicit calculation of $f_{ij}(\bx(t_{c+1})\mid\bx(t_{c}), \cTc) $, the calculation of the overall likelihood via the Forward Algorithm can be simplified from the version in \S\ref{InCH} and \S\ref{sec:hetero}. We can regard the process as a heterogeneous \HMM\ just at the actual observation times $t_c$, but with observation probabilities given by equation (\ref{SpatHomMeth}) and transition probabilities 
\[
\pi_{ij}(c) = \sum_{\bs} \pi_{ij}(\bs).
\]
The sampling of the potential switching times on on $(t_a,t_b)$ uses a Poisson process of rate $\kappa$. 
Since no additional locations need to be sampled, we have only the actual locations $\Xobs$, and we have 
\begin{align*}
p(\newT\mid\fixT,\Xobs,\Theta) &\propto p(\Xobs|\newT,\fixT,\Theta)p(\newT|\Theta),\\
q(\newT\mid\oldT,\fixT,\Xobs,\Theta) &= q(\newT\mid\Theta)
\end{align*} 
and hence Hastings ratio
\[
\frac{p(\Xobs\mid\newT,\fixT,\Theta)}{p(\Xobs\mid\oldT,\fixT,\Theta)},
\]
calculated simply from two passes of the Forward Algorithm.

As in the heterogeneous case, the other steps which sample the model parameters are straightforward, using the Forward Algorithm to calculate the likelihood.

\subsection{Sampling the states}
\label{FFBS}
The key strength of the InCH approach is that likelihood evaluation, and therefore parameter estimation, does not need sampling of the behavioural states but instead integrates them out. However, we are often interested in reconstructing the states too, as part of the detailed interpretation of the the data and of the states themselves. In a standard \HMM\ setting, the Viterbi algorithm is typically used to calculate the single most likely state sequence, after the Forward Algorithm has been run to calculate the overall likelihood of the observations, and the same approach could be taken here. However, within the fully Bayesian approach, it seems more natural to \emph{sample} the hidden state sequence, if it is of interest. This can be achieved using the Forward-Filtering Backward-Sampling of \citet{Fruwirth}, originally designed for discrete-time models, adapted to the continuous-time context by making use of uniformization and by allowing the model to be time-heterogeneous. \citet{RaoTeh} use a similar idea as a key part of their \MCMC\ algorithm for a continuous-time Markov chain, giving a Gibbs sampler for the state sequence over some sub-interval within their data. See their Appendix A for details; adaptation for the current context is straightforward.  
This is of most interest in a more applied setting, but a simple example is given in \S \ref{backsample}.

\section{Implementation}
\label{sec:implement}
In each of the examples below, I chose to fix $\kappa$ so that $\kappa\delta t=1$ for the typical interval between observations. This ensures that for such an interval, the probabilities of 0, 1 or 2 potential switches are not too small (approximately 0.368, 0.368, 0.184 respectively), permitting visits to a behaviour to have a chance of being represented even if they do not span an observation. 

All runs were carried out on the same low-specification desk-top PC (2.90 GHz, 8.0GB). Runs were 100,000 iterations for the homogeneous model, or 200,000 iterations for the heterogeneous model, with burn-in of 10,000 iterations (exept where noted) and thinning by a factor of 100. The various tuning parameters, such as proposal variances for the Metropolis-Hastings steps, were optimized after Latin hypercube sampling, with 5 replicates at each sampled point. 

Coding is in R \citep{R3.4}, for ease of development, and there is in all cases scope for substantial improvement by re-writing in a fully compiled language. Relative speeds are therefore much more informative than absolute speeds.

Effective sample size was calculated using the package Coda \citep{CODA}, minimizing over the key unknown quantities. These include the model 
parameters and, where applicable, over the sampled behavioural states at the times of observations. They also include the counts of potential switching times in each interval between observations, as a proxy for the switching times themselves which would be less straightforward to assess. This minimum effective sample size was compared with the running time required, to give an effective sample size per second.

\section{Example: heterogeneous case}
\label{sec:ibex}
\subsection{Data}
As a small-scale example, I use irregular data consisting of 71 GPS relocations at approximately 4-hour intervals of an ibex in the Belledonne mountain in the French Alps, originating with the French 
\textit{Office national de la chasse et de la faune sauvage}
and made available in the ADEhabitatLT package \citep{ADEhabitat} for R. The majority of the intervals between observations were around 4 hours, plus or minus 90 seconds, but there were some `missing values' leading to eight intervals of around 8 hours, one of 12 hours, and one of 16 hours. 

\subsection{Model}

The model fitted was a two-state switching diffusion based on a division of the space into two regions, inside and outside a circular boundary. The boundary is intended as a simple representation of the animal's home range; its behaviour switches at some finite rate to `match' its location, inside or outside the boundary. The movement processes for the two states are Ornstein-Uhlenbeck processes \citep{Dunn1977,Blackwell1997,Exact} with a common centre of attraction. The boundary is taken to be known and fixed; this gives an adaptive model, in the sense of \citet{Exact}. Estimation of the boundary is possible within the \MCMC\ part of the algorithm; it is omitted here for simplicity, but is addressed
in an unpublished report by S.\ Tishkovskaya and P.\ G.\ Blackwell, available from the author, and also in a 2019 University of Sheffield School of Mathematics and Statistics PhD thesis by H.\ Alkhezi.

\subsection{Comparison of methods}

This adaptive model can be fitted exactly using the algorithm of \citet{Exact}, sampling potential switching times, corresponding locations, and the full behavioural trajectory. It can also be fitted using the InCH approach introduced in \S\ref{sec:hetero}, obviating the need to sample the behaviours. In each case I took $\kappa=0.25$. 

The algorithm of \citet{Exact} mixes rather poorly, because the space to be explored by the \MCMC\ algorithm includes the complete state trajectory at the potential switching times, behaviours as well as locations. The runs reported here, five replicates of $200,000$ iterations as described in \S\ref{sec:implement}, give effective samples sizes in some cases too small for definitive analysis, but sufficient for comparison with the new approach.

\subsection{Results: real data}
\label{sec:het-real}

Fitting the two-state adaptive model with fixed boundary, using the state-sampling algorithm of \citet{Exact}, 
five replicates of $2\times10^5$ iterations with the optimal tuning parameters took 
2240s. A burn-in of 50,000 iterations was required, and gave an effective sample size of 35.0, equivalent to 0.0156 samples per second.
Using the InCH approach, the corresponding running time was 6240s, for an effective sample size of 274, giving 0.0439/s.
\begin{figure}[!tb]
\includegraphics[scale=0.56]{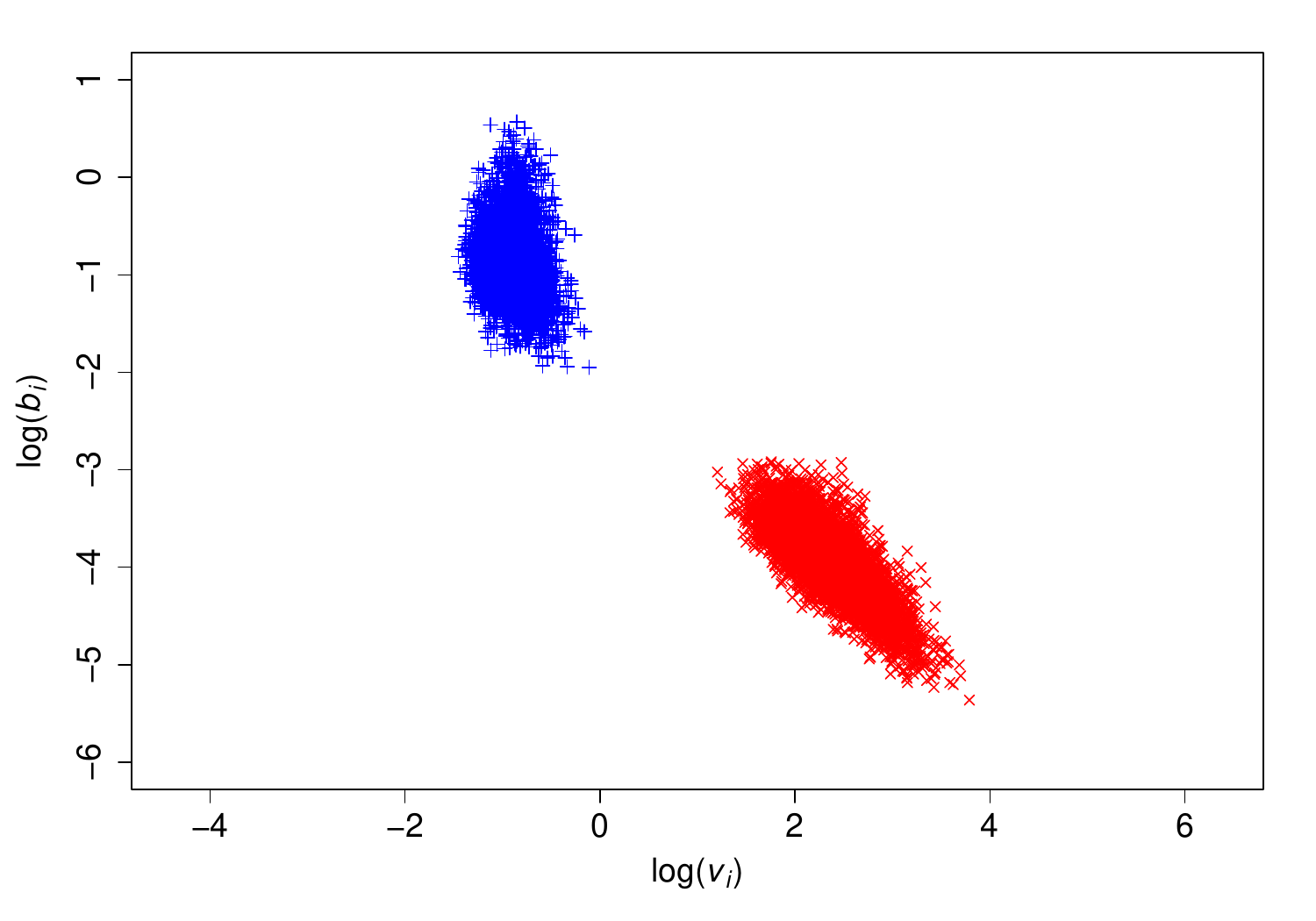}
\caption{Posterior distributions of the movement parameters for the two states in the heterogeneous model}
\label{fig_new_het_real}
\end{figure}
Posterior distributions of the movement parameters are given in Figure \ref{fig_new_het_real}.
In this small heterogeneous example, the InCH approach is around 2.8 times as efficient as the original `exact' algorithm, and of course remains exact in the same sense.

\subsection{Results: simulated data}
\label{sec:het-sim}

Using rounded versions of the point estimates from the analysis in \S\ref{sec:het-real}, I simulated a larger data-set of 201 observations at 4 hour intervals (so approximately three times the size of the real data). I analysed these simulated data in the same way as before, re-running the Latin hypercube tuning since a larger data-set changes the trade-off between running time, acceptance rate and mixing.

The method of \cite{Exact} took 2240s for an effective sample size of only 11.1, giving a sampling rate of 0.00493 per second. 
The InCH approach took 6780s for an effective sample size of 443, giving a sampling rate of 0.0653 per second.
%
\begin{figure}[!tb]
\includegraphics[scale=0.56]{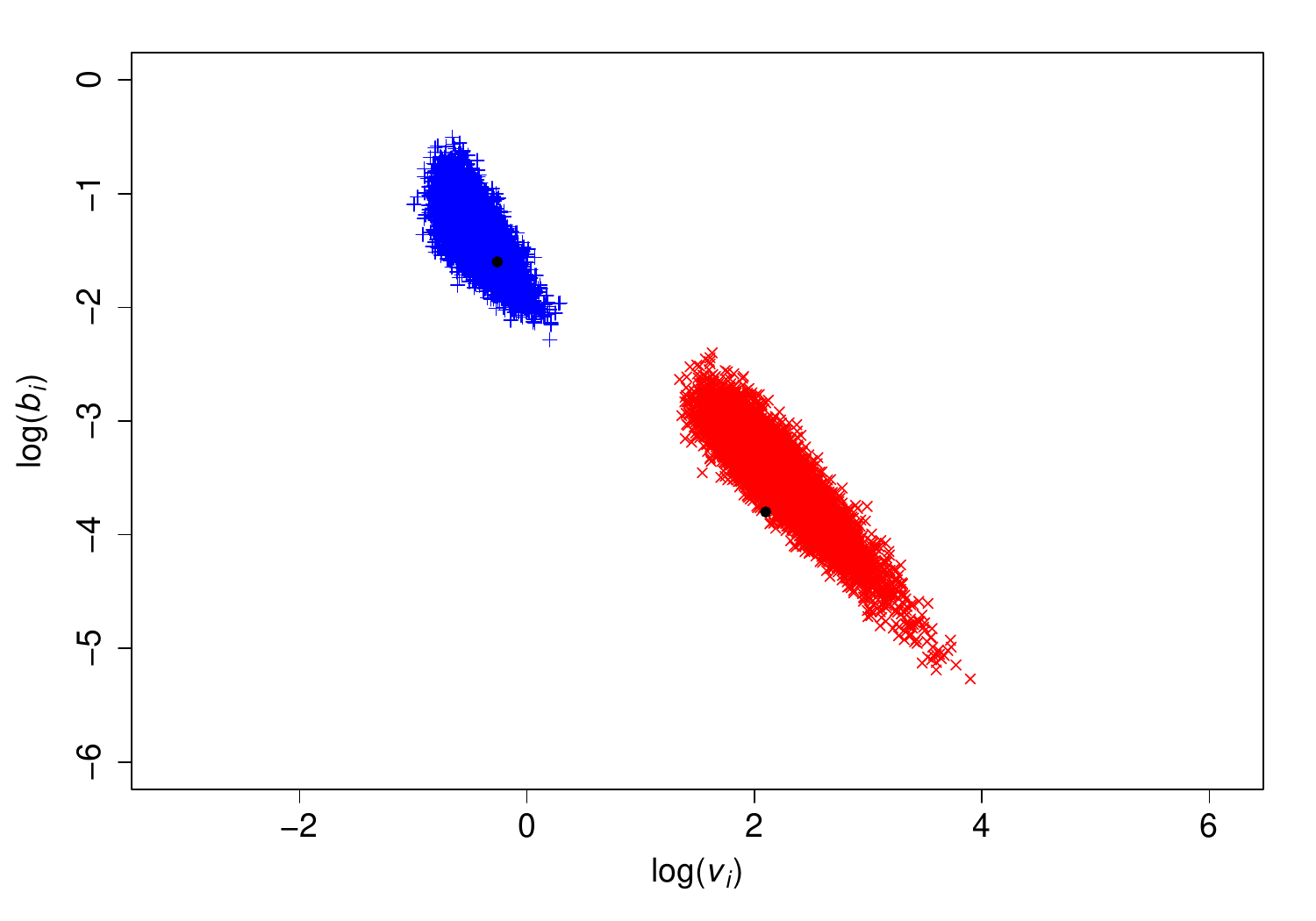}
\caption{Posterior distributions of the movement parameters for the two states in the heterogeneous model. True values are shown as solid dots.}
\label{fig_new_het_sim}
\end{figure}
Posterior distributions of the movement parameters are given in Figure \ref{fig_new_het_sim}.
These results are better than might be expected in the light of the previous example, particularly with the InCH method where the time taken per effective sample increases noticeably more slowly than the amount of data. Initial investigations suggest that this is due to the increased regularity in the simulated data, which for simplicity were simulated at regular intervals within no missing data. The few larger intervals in the real data impose a disproportionate computational cost, in both algorithms.

For this case, the InCH approach is around 13 times as efficient as the original algorithm.

\section{Example: homogeneous case}
\label{sec:kinkajou}

\subsection{Data}
\label{kink_intro}
As an example, I consider a small data-set of two-dimensional GPS locations for a kinkajou (\textit{Potos flavus}), taken from the Movebank Repository \citep{kinkajou_data}.
The data-set consists of 61 fixes, mostly of intervals of 9 and 11 minutes but with a few missing values leading to intervals of 20 to 30 minutes. 

\subsection{Model}
The model fitted is a simple InCH process, with $n$ different states each involving isotropic Brownian motion on $\mathbb{R}^2$ with a different speed (i.e.\ diffusion parameter) $v_l, l=1,\ldots,n$, in increasing order to avoid label switching. 

\label{BM}
In this case, the calculation of transition densities between observations, given in \S\ref{SpatHom}, has a particularly simple form. For particular states $i$ and $j$ at times $t_c$ and $t_{c+1}$, 
$\phi_{ij}(\bX(t_{c+1})\mid \bX(t_{c}),\cTc,\bs)$ in the notation of \S\ref{SpatHomMeth} is specified by
\[
\bX(t_{c+1})\mid \bX(t_{c}),\cTc,\bs \sim \text{N}(\mathbf{0}, v_{ij}(\bX(t_{c}),\cTc,\bs) I_2)
\]
where
\[
v_{ij}(\bX(t_{c}),\cTc,\bs) 
= (t_{c,1}-t_c)v_i + \sum_{k=1}^{M_c-1} (t_{c,k+1}-t_{c,k})v_{s_k} + (t_{c+1}-t_{c,m_c})v_j,
\]
that is, the appropriately time-weighted average of the diffusion parameters in different states, 
and $I_2$ is the $2\times2$ identity matrix.

The runs reported here have $n=3$, except the run showing the sampling of the states which has $n=2$ for ease of visualization. 

\subsection{Comparison of methods}

The model being fitted is spatially homogeneous, so the method of \S\ref{SpatHom} is appropriate here. For comparison, the same model could be fitted using the forward simulation method of \citet{Exact}, which is the origin of the thinned Poisson representation. The method there does not exploit \HMM\ computational methods, but instead tracks the whole state trajectory as part of its \MCMC\ algorithm, re-sampling a part of it at each iteration. Since the `full data' likelihood conditional on a complete trajectory for the states can be calculated more easily than the `integrated' likelihood of \S\ref{SpatHom}, the algorithm of \citet{Exact} runs more quickly.
However, it mixes much less well, because of the high dimension of the space to be explored by the \MCMC\ algorithm.
This comparison is arguably unfair, however, as the existing algorithm does not exploit the spatial homogeneity of the model or the simplification described in \S\ref{BM}. Instead, therefore, performance is shown for a version in which the behavioural sequence is sampled, and the simplification of \S\ref{BM} is applied. This is essentially the algorithm applied by \citet{Blackwell2003}, in the particular case where all movement processes are versions of Brownian motion, representing the state of the art for an exact analysis in the spatially homogeneous case, and therefore a fairer comparison.

\subsection{Results: real data}
\label{sec:real}
Firstly, results are shown for the analysis of the small data-set of 61 observations. The results support a 3-state model as being plausible, but the key point of interest here is computational performance, compared for the existing and proposed algorithms.

\begin{figure}[!htb]
\includegraphics[scale=0.56]{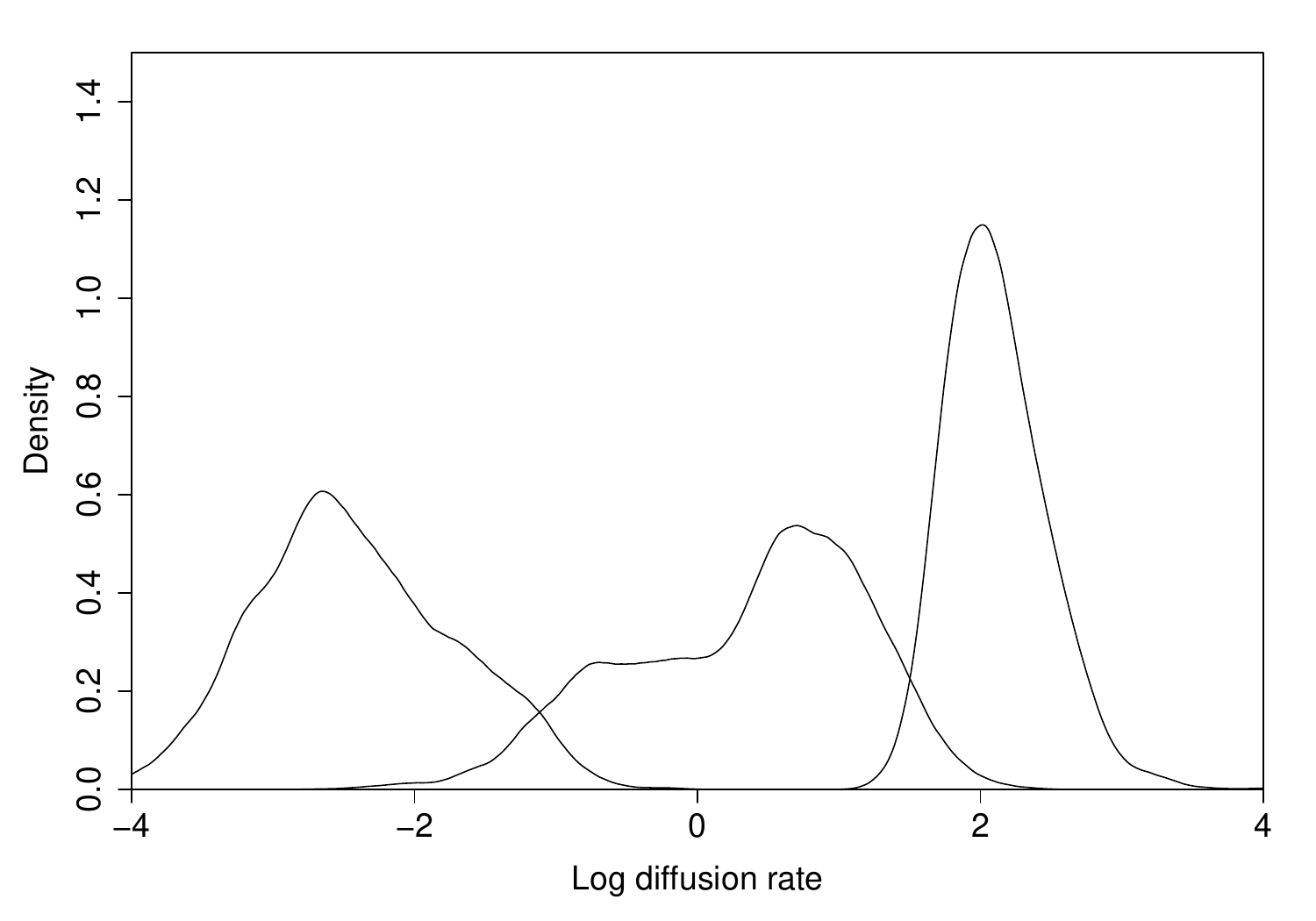}
\caption{Posterior distributions of the diffusion parameters for the three states in the homogeneous model.}
\label{fig_new_hom_real}
\end{figure}

Using the homogeneous version of the existing algorithm, as described above, 
five replicates of $10^5$ iterations took 682s. The effective sample size was 406, giving 
0.595 samples per second.  
Regarding the model as an InCH and using the Forward Algorithm to calculate the likelihood, five replicates of $10^5$ iterations took 1130s and produced an effective sample size of 569, giving a rate of 0.504/s. 
Posterior distributions of the movement parameters are given in Figure \ref{fig_new_hom_real}.
For this small sample, with a homogeneous model, the InCH approach is marginally less efficient than the existing approach.

\subsection{Results: simulated data}
The real strength of the InCH approach is that its efficiency scales better with the size of the data-set than existing methods. To illustrate this, I simulated data from the estimated parameters in \S\ref{sec:real} to obtain 
301 observations (i.e.\ simulating an observation period 5 times longer than the data used in \S\ref{sec:real}) and then analysed them in the same way as before. 
The values of the tuning parameters differ between the two data-sets, based on separate Latin hypercube optimization.

\begin{figure}[!htb]
\includegraphics[scale=0.56]{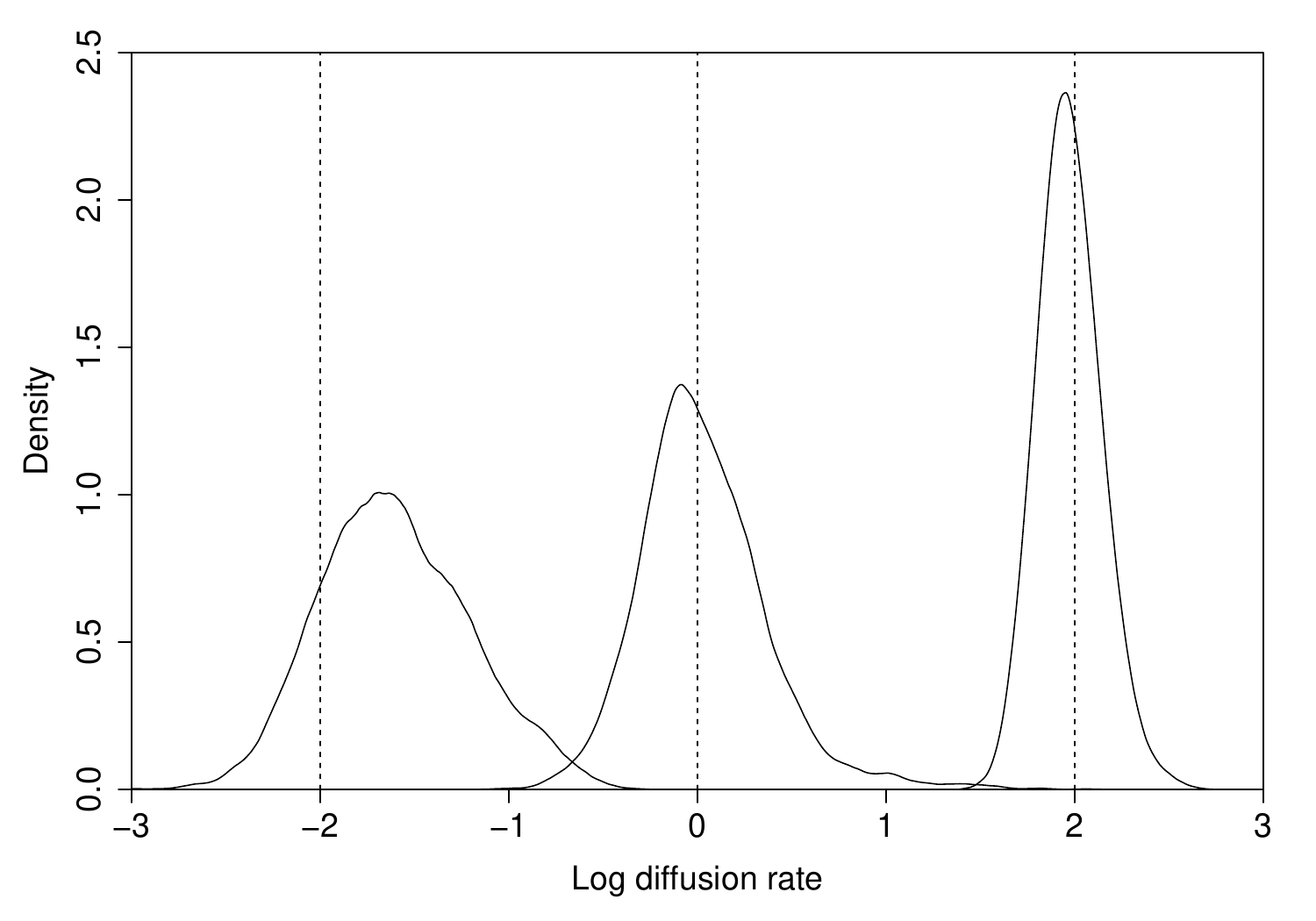}
\caption{Posterior distributions of the diffusion parameters for the three states in the homogeneous model.
True values are shown as vertical dashed lines.}
\label{fig_new_hom_sim}
\end{figure}

With the existing algorithm, five replicates of $10^5$ iterations took 1219s, so the running time is only around 1.8 times as long as for the smaller data-set. This is because much of the computational effort goes on localized updates to the behavioural sequence, for which only part of the likelihood needs to be evaluated. 
However, the effective sample size is only 26.6, 
giving a sampling rate of 0.0219/s. 
The InCH approach ran five replicates of $10^5$ iterations in 4340s, about 3.8 times as long as for the small data-set. It gave an effective sample size of 345, decreasing much more slowly than was the case for the existing method. This is because the dimension of the space over which the \MCMC\ algorithm is sampling is not increasing with size of the data-set, so mixing does not degrade so quickly. It does still decrease to some extent, because the optimal proportion of the Poisson $\kappa$ process to resample at each iteration (estimated through the Latin hypercube experiments) is decreasing. The net rate of generating independent samples is 0.0796/s, so for this data-set the InCH approach is around 3.6 times as efficient as the existing method.
Posterior distributions of the movement parameters are given in Figure \ref{fig_new_hom_sim}.

\subsection{State reconstruction}
\label{backsample}
\begin{figure}[htb]
\includegraphics[scale=0.56]{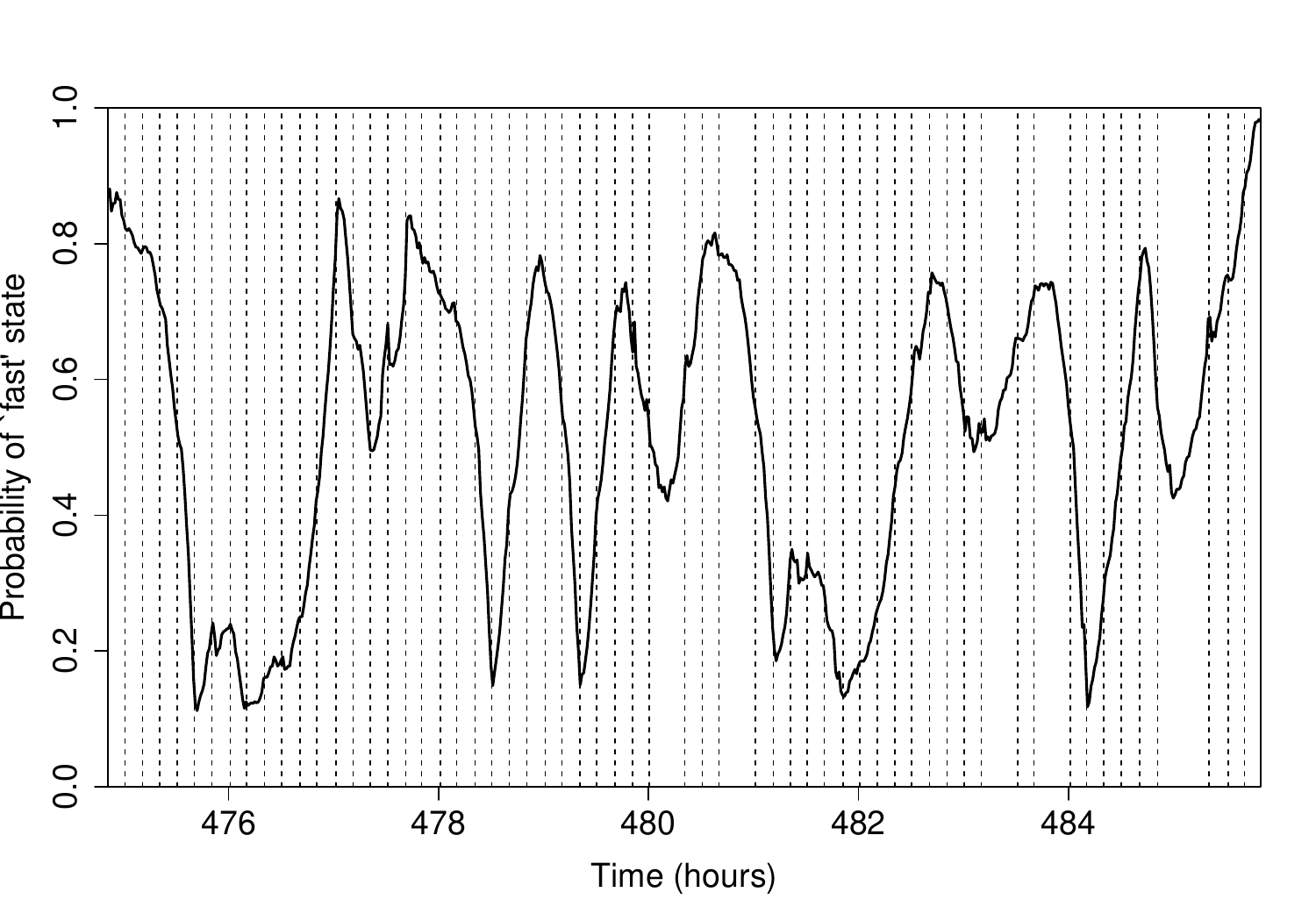}
\caption{Posterior probability that the kinkajou is in the `fast' state in the 2-state model (solid line). Times of observations (relative to an arbitrary origin) are indicated by the vertical dashed lines.}
\label{fig:decode}
\end{figure}

As described in \S \ref{FFBS}, reconstruction of the actual states is not necessary for parameter estimation, but may be of interest. For clarity, I have fitted a simpler model to the data of \S \ref{kink_intro}, of the same form as above but with only two states, and the results of sampling the behavioural states are summarized in Figure \ref{fig:decode}.
Based on a sample of size 5000 from the posterior distribution of states, obtained using the Forward-Filtering Backward-Sampling algorithm, posterior probabilities of being in the faster of the two states are calculated every minute during the time covered by the data. (Calculating the probability at every potential switching time, over all 5000 trajectories, would be prohibitively expensive, and would add essentially no further information.) Figure \ref{fig:decode} shows these posterior probabilities, and also indicates the times of the observations. The probabilities are often more extreme (closer to 0 or 1) just at the times of observations, and often close to 0.5 when there are missing data, as might be expected.

\section{Discussion}
\label{discuss}

\subsection{Summary of results}

I have shown that, while standard \HMM\ techniques do not apply directly to continuous-time movement models, a very broad class of such models can be be seen as \HMMs\ after conditioning on the Poisson process of potential times of behavioral change. This can be exploited within \anMCMC\ algorithm as a highly efficient way of evaluating likelihood for these models without sampling the behavioural states, resulting in much improved mixing. 
In an example, the scaling of computational performance with the size of data-set is shown to be much better in the new approach than in existing methods. Thus, it is possible to extend the key benefit of the \HMM\ approach to realistic continuous-time models.

\subsection{Extensions}

For definiteness of exposition, the models and algorithms above make a number of assumptions that are not in fact essential. 

I have assumed that behaviour is not observed at all, which is the most common case, though \citet{Blackwell2003} addresses the opposite case. Increasingly, partial information about behaviour is available, either through direct observation or through other kinds of telemetry such as accelerometry. The methods above can incorporate this extra information readily, by adding an extra term in the calculation of the likelihood at the time of the observation.

As is widespread in movement analysis, including discrete-time \HMMs, I have neglected GPS measurement error above. However, it can be readily incorporated by including extra variables in the state of the Markov chain used for inference, representing the true, rather than observed, location at the time of each GPS fix. See \citet{AlisonBAYSM} for an illustration of this in a similar context. Depending on the model, a more sophisticated Kalman filtering approach may also be possible c.f.\ \citet{TheoVelocity}.

As mentioned above, it is conceptually simplest to keep the rate $\kappa$ of potential switches as a constant. However, that requires the prior distributions for the rates $\lambda_i(\cdot)$ to be bounded above. An alternative is to allow $\kappa$ to be data-driven, via the $\lambda_{ij}(\cdot)$s. Some care is needed, since $\kappa$ is not really a parameter but rather a computational device (for example, increasing $\kappa$ does not change the model at all, though it slows the calculation), but 
one successful approach is explored in a 2019 University of Sheffield School of Mathematics and Statistics PhD thesis by H.\ Alkhezi.

Finally, it may be desirable to allow behavioural switching to depend on the length of time already spent in the current state, as well as the absolute time and other covariates, as a kind of semi-Markov extension. This can be done readily either by simply incorporating this elapsed time as an argument to $\lambda_{ij}(\cdot)$, which complicates the computation somewhat, or by extending the state space; again, see the thesis by Alkhezi 
for details.

\section*{Acknowledgements}
I am grateful for comments, discussion, encouragement and questions from 
Hajar Alkhezi,
Fay Frost,
Ruth King, 
Roland Langrock,
Th\'eo Michelot,
Jordan Milner,
Alison Poulston,
Toby Patterson
and
Len Thomas.

\bibliographystyle{biometrika}
\bibliography{../../papers/PGB,../../papers/move,../../papers/AStAreview,../../CVetcSelected/PGBpapers_only_unaccepted}
\end{document}